# Uniting the wave and the particle in quantum mechanics

**Peter Holland**[1]

**Abstract**

We present a unified field theory of wave and particle in quantum mechanics. This emerges from an investigation of three weaknesses in the de Broglie-Bohm theory: its reliance on the quantum probability formula to justify the particle guidance equation; its insouciance regarding the absence of reciprocal action of the particle on the guiding wavefunction; and its lack of a unified model to represent its inseparable components. Following the author's previous work, these problems are examined within an analytical framework by requiring that the wave-particle composite exhibits no observable differences with a quantum system. This scheme is implemented by appealing to symmetries (global gauge and spacetime translations) and imposing equality of the corresponding conserved Noether densities (matter, energy and momentum) with their Schrödinger counterparts. In conjunction with the condition of time reversal covariance this implies the de Broglie-Bohm law for the particle where the quantum potential mediates the wave-particle interaction (we also show how the time reversal assumption may be replaced by a statistical condition). The method clarifies the nature of the composite's mass, and its energy and momentum conservation laws. Our principal result is the unification of the Schrödinger equation and the de Broglie-Bohm law in a single inhomogeneous equation whose solution amalgamates the wavefunction and a singular soliton model of the particle in a unified spacetime field. The wavefunction suffers no reaction from the particle since it is the homogeneous part of the unified field to whose source the particle contributes via the quantum potential. The theory is extended to many-body systems. We review de Broglie's objections to the pilot-wave theory and suggest that our field-theoretic description provides a realization of his hitherto unfulfilled 'double solution' programme. A revised set of postulates for the de Broglie-Bohm theory is proposed in which the unified field is taken as the basic descriptive element of a physical system.

> *...our model in which wave and particle are regarded as basically different entities, which interact in a way that is not essential to their modes of being, does not seem very plausible. The fact that wave and particle are never found separately suggests that they are both different aspects of some fundamentally new kind of entity...*
>
> David Bohm [1]

## 1 Introduction

The supposition that the wavefunction $\psi$ provides the most complete description of the state of a physical system that is in principle possible has permeated interpretational discourse since quantum theory's inception. Yet it is an arbitrary and unproven conjecture adopted out of theoretical choice rather than empirical imperative. Indeed, the almost boundless paradoxical and fantastical narratives that flow from it dissolve when enhanced concepts of the quantum state are admitted. The most successful example of such a completion is the de Broglie-Bohm causal interpretation, or pilot-wave theory, where in addition to $\psi$, conceived as a physically real guiding field, the state comprises a material corpuscle traversing a well-defined spacetime trajectory. Rather than pertaining to mutually exclusive experiments, 'wave-particle duality' becomes an objective and permanent feature of matter, and all within 'one world'.

    Whilst the virtues of the de Broglie-Bohm theory cannot nowadays be gainsaid, its standard exposition is open to question. In the next section, we identify three areas where the theory would benefit from theoretical development, relating to the justification of its postulates and how the wave-particle interaction is modelled, and sketch how these problems may be addressed by founding the theory on analytical principles. The remainder of the paper works out

---

[1] Green Templeton College, University of Oxford, peter.holland@gtc.ox.ac.uk

this programme in detail and culminates in a proposed reformulation of the de Broglie-Bohm theory based on a unified field whose law of motion combines those of the field and the particle.

## 2 Critique of the theory of de Broglie and Bohm

### *2.1 The postulates*

For a single-body system the de Broglie-Bohm theory is usually presented in terms of the following three postulates governing the system's wave and particle constituents ($i,j,k,\ldots = 1,2,3$; $\partial_i = \partial/\partial x_i$):

1 The wave is described mathematically by the wavefunction $\psi(x,t)$ obeying the Schrödinger equation

$$i\hbar \frac{\partial \psi}{\partial t} = -\frac{\hbar^2}{2m} \partial_{ii} \psi + V\psi. \tag{1.1}$$

2 Writing $\psi = \sqrt{\rho} \exp(iS/\hbar)$, the particle is a point obeying the guidance equation

$$m\dot{q}_i = \partial_i S(x,t)\big|_{x=q(t,q_0)} \tag{1.2}$$

whose solution $q_i(t,q_0)$ depends on the initial coordinates $q_{0i}$. The latter coordinate any space point where the initial wavefunction $\psi_0(x) \neq 0$ and they identify the trajectory uniquely.

3 For an ensemble of wave-particle systems with a common $\psi$ component, the particle spatial probability density at time $t$ is $\rho(x,t)$.

The theory is easily extended to an *n*-body system where the configuration space wavefunction is accompanied by *n* corpuscles moving in three-dimensional space [2].
    These are the bare bones of the theory although other concepts, such as energy and force, play a fundamental role [2]. Might the postulates be replaced by others that are perhaps better motivated? In examining this question, we start by exhibiting two simple ways in which the postulates may be usefully reformulated, relating to their interdependence and conceptual flexibility. These reformulations will be useful later. In this endeavour it is helpful to write the Schrödinger equation (1.1) as two real equations, valid where $\psi \neq 0$:

$$\frac{\partial \rho}{\partial t} + \frac{1}{m} \partial_i (\rho \partial_i S) = 0 \tag{1.3}$$

$$\frac{\partial S}{\partial t} + \frac{1}{2m} \partial_i S \partial_i S + Q + V = 0 \tag{1.4}$$

where $Q(x,t) = -(\hbar^2/2m\sqrt{\rho}) \partial_{ii} \sqrt{\rho}$ is the quantum potential.
    In the first reformulation, we show that postulates 1 and 2 imply that postulate 3 need be asserted only at one time. This follows on using (1.2) to write (1.3) equivalently as the local law of conservation of probability:

$$\frac{d}{dt}\left[\rho(q(t),t)d^3q(t)\right]=0 \tag{1.5}$$

where $d/dt = \partial/\partial t + \dot{q}_i \partial_i$. Then

$$\rho(q(t),t)d^3q(t) = \rho_0(q_0)d^3q_0 \tag{1.6}$$

so that, given the trajectory, $\rho$ is implied by $\rho_0$: $\rho(x,t) = \det(\partial q/\partial q_0)\rho_0(q_0)$ with $q_{0i} = q_{0i}(x,t)$. It follows that postulate 3 may be replaced without loss of generality by the postualte[2]

3* For an ensemble of wave-particle systems with a common $\psi$ component, the initial spatial particle probability density is $\rho_0(x)$.

In the second reformulation, we show how the first-order ('Aristotelian' [3]) law of postulate 2 may be expressed equivalently as a second-order ('Newtonian') equation, which brings out the fundamental role played by force in the causal explanation. In the context of the second-order equation the status of the first-order law is that it constrains the initial velocity of the particle, the constraint being preserved by the second-order equation. This result may be formulated as follows [4]:

***Proposition 1 (equivalence of first- and second-order laws)*** Let $m\dot{q}_i = \partial_i S_0(q)$ at $t = 0$. Then, for all $t$, $m\dot{q}_i = \partial_i S(x,t)\big|_{x=q(t)}$ if and only if $m\ddot{q}_i = -\partial_i(V(x,t)+Q(x,t))\big|_{x=q(t)}$. (Proof: Appendix A)

It is easy to show that the second-order equation also preserves the single-valuedness condition $\oint \partial_i S\, dx_i = nh$ and hence this need only be postulated at $t = 0$ [5].

Evidently, the first-order version of the dynamics is no less contingent than the second-order one since both hinge on the initial constraint; if that is disobeyed, the first-order version fails completely and the second-order version admits 'too many' solutions, i.e., trajectories that in general do not preserve the probability (1.6) and hence may violate postulate 3, even if postulate 3* is obeyed (for a more general discussion of this point based on a quantum Liouville equation see [6]). These remarks reinforce the falsity of regarding one of the equivalent formulations of the guidance law as more 'fundamental' than the other. Indeed, understanding how the 'piloting' or 'guidance' of the particle comes about, epithets that are commonly assigned to the first-order version, necessitates invoking the second-order force law. It is, of course, reasonable, in a theory whose remit is to counteract the vagueness of conventional interpretations, to utilize the full set of concepts that it is able to make precise. For the theory is not about the particle trajectory *per se* but about why it *changes*, so that the configuration of matter at one instant is causally and continuously connected to its later and preceding configurations. The limited explanatory power of the first-order law is apparent in many examples. For example, in stationary states where the particle or system of particles may be in uniform motion, it is the quantum force that explains phenomena such as the stability of the atom, the attraction of neutral atoms to form a molecule, the Casimir effect, and the pressure exerted by a gas [2,6].

---

[2] There is as yet no entirely satisfactory justification for this postulate within the de Broglie-Bohm theory. But in this regard the theory is in no worse shape than conventional quantum mechanics where the probability formula is likewise postulated.

## 2.2 Limitations

A weakness of the stated postulates is that the choice of guidance law in postulate 2 is dictated, in the first instance, by compatibility with postulate 3, i.e., it is justified by the statistics it conserves. But a wide class of non-trivially distinct guidance laws is compatible with a conserved $\rho$-distribution, if that is the only selection criterion [7]. This criterion is akin in classical mechanics to attempting to justify Newton's law for the motion of an individual by inference from Liouville's equation describing the evolution of an ensemble. Moreover, conclusions drawn from the particle law beyond its remit of conserving the quantum probability may be suspect. For example, if we wish to dispense with postulate 3 and attempt to derive it using an argument based on the guidance law, as first suggested by Bohm [3], we risk circularity if that postulate has already been invoked to justify the law. Another example is the use of the trajectory law to compute quantities that go beyond the standard quantum formalism, such as transit time [8].

Hence, it is desirable to find a justification for the guidance law by studying the dynamics of an individual wave-particle composite. But here we encounter two further shortcomings of the customary model, stemming from its somewhat primitive representation of wave-particle duality.

The first issue concerns what is arguably the model's most striking aspect (although historically one of its least analyzed): the tacit assumption of postulates 1 and 2 that the particle responds passively to the wave without reciprocal action. The absence of particle reaction is not a logical problem; indeed, it is crucial if one aims to avoid disturbing the Schrödinger evolution and hence maintain the usual predictions of quantum mechanics (which of course are independent of the corpuscle). Nevertheless, it is a singular occurrence in physics that solicits scrutiny. To set it in context, and bearing in mind the hydrodynamic analogy in quantum mechanics [2], the no-reaction assumption is comparable to introducing a tracer into a classical fluid and assuming that it will follow a streamline, i.e., adopt the local fluid velocity along its route. In fact, such a hypothesis needs careful examination of the internal structures and mutual actions of the dirigible particle and the fluid, according to the dynamical principles expected to govern such interactions (Newton's laws in that case, e.g., [9]). Likewise, one may seek a suitable dynamical framework for the quantum wave-particle interaction within which the absence of reaction is a natural attribute rather than an incidental oddity. Of course, the details of such a theory need not mirror classical treatments.

The second deficiency of the wave-particle model is that, as remarked by de Broglie (see Appendix B) and Bohm [1], the innate dissimilarity of the point-particle and field enlisted as constituents of the composite is belied by the core assumption of the model that these entities are inseparable. There is nothing in their natures that would dictate that one cannot be present without the other. Can the two elements be connected at a more basic level so that their intimacy becomes an aspect of a unified description?

To summarize, we have identified three avenues of enquirty where improvements to the conventional presentation of the de Broglie-Bohm theory may be usefully sought *viz*. detaching the dynamical law of the particle from probability, finding a theoretical environment to represent the absence of particle reaction, and developing a harmonious model where the wave and particle become aspects of a unified structure.

## 2.3 Alternative approach

We shall show that the clutch of issues just mentioned may be addressed, together, within an analytical theory of the wave-particle interaction. This approach develops previous work devoted to the first two problems, i.e., finding a non-statistical justification for the particle law and a theoretical context to represent no-reaction [6,10]. In this work it was assumed that the wave's action on the particle is mediated by a scalar potential, and that the composite exhibits no observable differences with a quantum system. Thus, when the composite is the 'source' of another system – of an electromagnetic field through the current it generates, for instance, or of a gravitational field through its energy-momentum complex (regarded as the low-energy limit of a relativistic tensor) – that system should not 'see' more than a 'quantum system', i.e., the composite should behave in this regard like the bare Schrödinger field. Specifically, we required that the conserved matter, energy and momentum densities of the composite coincide with their Schrödinger counterparts. On the basis of these constraints (and another, time reversal covariance, that was assumed tacitly) it was found that the particle variables obey the de Broglie-Bohm law while the interaction with the wave is mediated by the quantum potential. To our knowledge this was the first physical (as opposed to statistical or mathematical) justification for the de Broglie-Bohm law that does not involve invoking further processes, such as stochastic fluctuations.

The formulas for the number, energy and momentum densities of the composite employed in this scheme are unknown *a priori*. A natural way to obtain them is as consequences of symmetries the system may reasonably be expected to possess (specifically, global spacetime and gauge translations), for then we may employ Noether's theorem. This procedure also allows us to investigate the role of symmetries in establishing a consistent quantum particle theory, an issue that has been examined extensively in the case of Lorentz covariance but is otherwise underexplored.

The previous presentation was rather terse and no reference was made to the potential role of the analytical theory in addressing the third problem described above, that of finding a unified description. Our purpose here is to give a more thorough account of the analytical theory and bring out its implications for a unified theory.

To summarize, we develop the theory of a corpuscle interacting with a Schrödinger wave, conceived as a single physical system, using the following four assumptions:

1 The system may be treated using analytical methods.
2 The particle is acted upon by $\psi$ via a scalar potential without reaction.
3 The theory admits global gauge, space and time translations as symmetries.
4 The corresponding conserved matter, energy and momentum densities implied by Noether's theorem coincide with their Schrödinger values.

We shall see that these assumptions provide a consistent theory in which the particle law of motion and the interaction potential are fixed as functions of the wavefunction up to an undetermined constant. No further constraints are obtainable using the method of equal densities. The undetermined constant may be fixed using either of the following additional assumptions, each of which results in the de Broglie-Bohm law and the quantum potential as mediating potential:

5a The interaction potential is a scalar under time reversal.
5b For an ensemble of composite systems with a common $\psi$ component, the particle spatial probability density at time $t$ is $\rho(x,t)$.

Case 5a implies the guidance formula for an individual system (our original goal) and 5b shows how an alternative statistical derivation works in this context.

This method clarifies other issues associated with the de Broglie-Bohm theory: the nature of the mass of the composite system; and how the energy and momentum of the composite are conserved under the usual conditions on the external potential required in quantum mechanics.

Our principal results pertain to the novel unified perspective that emerges as a concomitant of the analytical technique. The absence of reciprocal action is incorporated since $\psi$ is represented as the homogeneous component of an inhomogeneous unified field and so is naturally source-free. Moreover, the unified field – to whose source the particle contributes – integrates into its structure both the $\psi$ field and the particle (as a highly concentrated amplitude), and its inhomogeneous wave equation unites the Schrödinger and particle-guidance equations. These features survive extension of the theory to many-body systems.

It turns out that this work bears a close affinity with de Broglie's putative 'double solution' theory, which we review in Appendix B. This programme was connected with de Broglie's critical analysis of the pilot-wave theory, a chronicle that seems not to be widely known. In fact, the problem of finding a matter equation of the type desired (but never given) by de Broglie was in effect solved by the author in the earlier work treating the back-reaction problem [6,10] but no reference was made to the double solution since the connection was not noticed.

### 3 Analytical derivation of the guidance equation for a wave-particle composite

#### *3.1 Euler-Lagrange equations for a corpuscle interacting with a Schrödinger wave*

We consider a system comprising a field whose state is described by the wavefunction $\psi(x,t)$ and a particle whose state is given by the Cartesian coordinates $q_i(t)$. This composite is our 'quantum system'. The particle moves under the influence of the field via, we shall assume, a scalar potential which is to be determined. The field obeys an unmodified Schrödinger equation due to the requirement that the particle does not react upon it.

For variational purposes the wavefunction $\psi$ and its complex conjugate $\psi^*$ are regarded as independent coordinates (equivalent to two real fields) and are varied independently. The variational procedure may be formulated as a constraint problem: a particle moves under the influence of fields constrained to obey Schrödinger's equation and its complex conjugate. A suitable Lagrangian is

$$L(x,q) = \tfrac{1}{2} m \dot{q}_i \dot{q}_i - V(q(t),t) - V_Q(q(t),t) + \int \left\{ \tfrac{1}{2} u^* \left[ i\hbar \dot{\psi} + (\hbar^2/2m) \partial_{ii} \psi - V\psi \right] + \text{cc} \right\} d^3 x. \qquad (2.1)$$

Here 'cc' means 'complex conjugate', $\dot{\psi} = \partial \psi(x,t)/\partial t$, $V$ is the classical external potential energy, and the potential energy $V_Q(\psi,\psi^*)$ represents the quantum effects on the particle. The complex functions $u(x,t)$ and $u^*(x,t)$ are independent Lagrange multipliers. The significance of the function $u$ will be explored below but its notation is chosen deliberately as it will be found to possess many of the key properties de Broglie demanded of the $u$ component of the 'double solution'.

It will prove convenient sometimes to use the polar variables $\rho$ and $S$ in place of $\psi$ and to mix these representations. It is assumed that $\psi \to 0$ as $x_i \to \pm\infty$. The action $\int L dt$ (and variants of it used below) is taken to be invariant with respect to global gauge, space and time translations. To ensure this, we make the following assumptions about the functions $V_Q$ and $u$:

(a) $V_Q$ depends on $x_i, t$ only implicitly through local functions of $\psi$, $\psi^*$ and their derivatives.

(b) $V_Q$ is a scalar with respect to independent global gauge, space and time translations (see Sec. 3.3), and so depends on $\psi$ and $\psi^*$ via $\rho$ and the space derivatives of $\rho$ and $S$ to any order (but not $S$ itself due to the gauge symmetry).

(c) $V_Q$ is independent of the time derivatives of $\psi$ and $\psi^*$. This is ensured if $V_Q$ is a scalar under Galilean boosts ($x'_i = x_i - w_i t$, $t' = t$) but we shall derive this property rather than posit it since it requires that $V_Q$ is free of $\partial_i S$ from the outset.

(d) $u$ transforms like $\psi$ under global gauge, space and time translations.

Variation of the action with respect to $u^*$ and $u$ yields the Schrödinger equation

$$i\hbar \frac{\partial \psi(x)}{\partial t} = -\frac{\hbar^2}{2m}\partial_{ii}\psi(x) + V(x)\psi(x) \qquad (2.2)$$

and its complex conjugate, respectively. Next, varying with respect to $\psi^*$ and $\psi$ gives the equation obeyed by the Lagrange multiplier,

$$i\hbar \frac{\partial u(x,t)}{\partial t} = -\frac{\hbar^2}{2m}\partial_{ii}u(x,t) + V(x)u(x,t) + 2\frac{\delta V_Q(\psi(q),\psi^*(q))}{\delta \psi^*(x)}\bigg|_{q=q(t)}, \qquad (2.3)$$

and the complex conjugate equation, respectively. Finally, varying the variables $q_i$ generates the particle equation

$$m\ddot{q}_i = -\frac{\partial}{\partial q_i}(V + V_Q)\bigg|_{q=q(t)}. \qquad (2.4)$$

The functional derivative in (2.3) is defined in Appendix C. The equations are completed by specifying the potential $V_Q$ and the initial conditions $\dot{q}_{0i}$, $q_{0i}$, $\psi_0$ and $u_0$. Expressions for $\dot{q}_{0i}$, $u_0$ and $V_Q$ will be obtained below from the consistency constraints imposed on the system. The only free variables in the system will then be $q_{0i}$ and $\psi_0$, on which $u$ will depend.

As required, the Schrödinger equation (2.2) is unmodified by the particle variables while the wavefunction appears in the particle law (2.4). To bring out the significance of the Lagrange multipliers $u$ and $u^*$, we note that they are the canonical momenta corresponding to $\psi^*$ and $\psi$, respectively (up to multiplicative constants):

$$p_{\psi^*} = \frac{\delta L}{\delta \dot{\psi}^*(x)} = -\frac{i\hbar}{2}u(x), \quad p_\psi = \frac{\delta L}{\delta \dot{\psi}(x)} = \frac{i\hbar}{2}u^*(x) \qquad (2.5)$$

The more significant property of the function $u$ here, however, is that its dynamical equation (2.3) is an inhomogeneous Schrödinger equation that exhibits on the right-hand side a (singular) source term that depends on $\psi$ and on $q_i(t)$ via $\delta(x-q(t))$ and its derivatives, which we shall examine below.

### 3.2 Fully field-theoretic formulation

To obtain local conserved quantities associated with the composite system using Noether's theorem it will be helpful to first write the particle and field variables appearing in the

Lagrangian (2.1) in a unified field-theoretic language. Consider the following Lagrangian density:

$$\mathcal{L}' = \left[\tfrac{1}{2}mv_i v_i - V(x) - V_Q(\psi(x,t))\right]\delta(x-q(t)) + \left\{\frac{1}{2}u^*\left[i\hbar\dot{\psi} + (\hbar^2/2m)\partial_{ii}\psi - V(x)\psi\right] + \text{cc}\right\} \quad (2.6)$$

obtained by a change of coordinates $q_i(q_0,t) \to x_i$ where the two sets of coordinates are connected by the relation

$$v_i(x,t) = \dot{q}_i(q_0,t)\big|_{q_0(x,t)}. \quad (2.7)$$

The integral of the density gives the Lagrangian (2.1): $L = \int \mathcal{L}' d^3x$. In this formulation the particle component of the dynamics is recast as a 'single-particle fluid dynamics' whose Eulerian picture is characterized by the density field $\rho_p(x,t) = \delta(x - q(q_0,t))$ and velocity field $v_i(x,t)$. $L$ now has a field-theoretic form but this is not suitable to obtain the particle law of motion (2.4) through a variational procedure because the time derivatives of the particle variables are missing, a well-known issue when passing from discrete to field descriptions. This problem may be tackled by introducing a Clebsch parameterization for the velocity [11] (the constant $m$ is introduced for later convenience):

$$v_i = m^{-1}(\partial_i\theta + \alpha\partial_i\beta). \quad (2.8)$$

Employing the new fields $\alpha(x,t), \beta(x,t)$ and $\theta(x,t)$, we shall use the following redefined Lagrangian density:

$$\mathcal{L} = -\rho_p(x,t)\left[\dot{\theta} + \alpha\dot{\beta} + (2m)^{-1}(\partial_i\theta + \alpha\partial_i\beta)^2 + V(x) + V_Q(\psi(x))\right]$$
$$+ \left\{\tfrac{1}{2}u^*\left[i\hbar\dot{\psi} + (\hbar^2/2m)\partial_i\partial_i\psi - V(x)\psi\right] + \text{cc}\right\}. \quad (2.9)$$

where $\dot{\theta} = \partial\theta(x,t)/\partial t$ etc. As desired, this gives a fully field-theoretic formulation of the interacting systems, with the particle variables represented by the fields $\rho_p$, $\alpha, \beta, \theta$ and the associated requisite time derivatives. The associated Lagrangian $L' = \int \mathcal{L} d^3x$ differs from $L$ in (2.1) by more than an added total time derivative so the associated action functionals are not the same: $\int L' dt \neq \int L dt$. Nevertheless, the revised set of Euler-Lagrange equations still give (2.2)-(2.4). The latter assertion is obvious for the variations of $\psi$, $\psi^*$, $u$ and $u^*$. We shall check explicitly that the particle equation (2.4) is obtained via the variations of $\alpha, \beta, \theta$ and $\rho_p$. We have

$$\delta\rho_p: \quad \dot{\theta} + \alpha\dot{\beta} + \tfrac{1}{2m}(\partial_i\theta + \alpha\partial_i\beta)^2 + V(x) + V_Q(x) = 0 \quad (2.10)$$

$$\delta\theta: \quad \frac{\partial\delta(x-q(t))}{\partial t} + \partial_i\left[\delta(x-q(t))v_i\right] = 0 \quad (2.11)$$

$$\delta\beta, \delta\alpha: \quad \dot{\alpha}+v_i\partial_i\alpha=0, \quad \dot{\beta}+v_i\partial_i\beta=0. \tag{2.12}$$

The last two equations follow upon using (2.11) which implies (2.7) using the identity (cf. Appendix G)

$$\frac{\partial\delta(x-q(t))}{\partial t}+\dot{q}_i\partial_i\left[\delta(x-q(t))\right]=0. \tag{2.13}$$

To deduce the particle law, we take the gradient of (2.10) and use the relations (2.8) and (2.12) to obtain an Euler-type force law:

$$m\left(\frac{\partial v_i}{\partial t}+v_j\partial_j v_i\right)=-\partial_i(V+V_Q). \tag{2.14}$$

In the field-theoretic picture the motion is characterized by the velocity field rather than the trajectory (as in the Eulerian picture in hydrodynamics). To obtain the trajectory version, we transform the independent variables $x_i \to q_i(q_0,t)$ via the relation (2.7) and the Euler law (2.14) becomes the particle equation (2.4) with initial condition $v_i(x,0)=\dot{q}_{0i}$. The 'equation of motion' (2.7) is essentially a conversion formula between the field and trajectory views of the particle motion. The equations (2.10) and (2.12) hold along each of the tracks $q_i(q_0,t)$ potentially occupied by the particle that are obtained by varying $q_{0i}$.

There is evidently some redundancy in the field-theoretic formalism for the particle component of the system since the variables $\alpha,\beta$ play no essential role. We shall remove these superfluous quantities by choosing the solutions $\alpha=\beta=0$ to (2.12). Then the velocity field reduces to $v_i=(1/m)\partial_i\theta$, the usual relation obtained in Hamilton-Jacobi theory for a structureless particle in an external scalar potential if the function $\theta$ is identified with the Hamilton-Jacobi function. That this is a correct identification follows since (2.10) reduces to the Hamilton-Jacobi equation. The field-theoretic version of the particle law of motion is therefore expressed by the equations

$$v_i(x,t)=m^{-1}\partial_i\theta, \quad \dot{\theta}+(2m)^{-1}\partial_i\theta\partial_i\theta+V+V_Q=0. \tag{2.15}$$

As shown above, these equations imply the particle law in the form (2.4) with initial conditions $\dot{q}_{0i}=m^{-1}\partial_i\theta_0(q_0)$ and $q_{0i}$ chosen freely.

Henceforth, we take (2.2), (2.3) and (2.15) as the dynamical equations for the composite system. The Lagrangian density is

$$\mathcal{L}=-\delta(x-q(t))\left[\dot{\theta}+(2m)^{-1}\partial_i\theta\partial_i\theta+V(x,t)+V_Q(\psi(x))\right] \\ +\left\{\tfrac{1}{2}u^*\left[i\hbar\dot{\psi}+(\hbar^2/2m)\partial_{ii}\psi-V(x)\psi\right]+\text{cc}\right\}. \tag{2.16}$$

### 3.3 Conserved densities obtained from Noether's theorem

We now apply Noether's theorem (see Appendix C) which asserts: to each continuous transformation that leaves the action $\int\mathcal{L}d^3x\,dt$ invariant there corresponds a density P and current density $P_i$ that together obey the continuity equation (C.7).

Using formula (C.5), the conserved number density corresponding to invariance of the action under an infinitesimal gauge transformation $\theta' = \theta + \varepsilon\eta$, $\psi' = \psi + i\varepsilon\eta\psi/\hbar$ (so that $S' = S + \varepsilon\eta$), $u' = u + i\varepsilon\eta u/\hbar$ with $\eta$ = constant and all other variables invariant is

$$J_0 = P/(-\eta) = \delta(x - q(t)) + \tfrac{1}{2}(u^*\psi + u\psi^*). \tag{2.17}$$

Although it does not assist in proving the guidance theorem, it is instructive to give the corresponding current density for the gauge symmetry obtained from (C.6):

$$J_i = P_i/(-\eta) = m^{-1}\partial_i\theta\delta(x - q(t)) + (i\hbar/4m)(\psi\partial_i u^* - u^*\partial_i\psi + u\partial_i\psi^* - \psi^*\partial_i u) - X_i \tag{2.18}$$

with

$$X_i = -\delta(x-q)\frac{\partial V_Q}{\partial(\partial_i S)} + \partial_j\left[\delta(x-q)\frac{\partial V_Q}{\partial(\partial_{ij} S)}\right] - \partial_{jk}\left[\delta(x-q)\frac{\partial V_Q}{\partial(\partial_{ijk} S)}\right] + \ldots. \tag{2.19}$$

Using (2.2), (2.3) and the easily proved result

$$\partial_i X_i = \frac{\delta V_Q}{\delta S} \tag{2.20}$$

it is readily confirmed that the number density and number current density obey the continuity equation (C.7):

$$\frac{\partial J_0}{\partial t} + \partial_i J_i = 0. \tag{2.21}$$

In fact, the point particle (delta function) and field contributions to the density and current density obey separate continuity equations, which reflects the invariance of the action under independent shifts in $\theta$ and $S$. The continuity equation obtained in this way for the particle alone, corresponding to a displacement in $\theta$, is the identity (2.11).

Next, the momentum density follows from invariance of the action under a space translation $x_i' = x_i + \varepsilon\xi_i$, $q_i' = q_i + \varepsilon\xi_i$ with $\xi_i$ = constant and all other variables invariant:

$$\mathfrak{P}_i = P/(\xi_i) = \partial_i\theta\delta(x - q(t)) + (i\hbar/2)(-u^*\partial_i\psi + u\partial_i\psi^*). \tag{2.22}$$

Finally, the energy density follows from invariance under a time translation $t' = t + \varepsilon\xi_0$ with $\xi_0$ = constant and all other variables invariant:

$$\mathfrak{H} = P/(-\xi_0)$$
$$= \delta(x - q(t))\left(\tfrac{1}{2m}(\partial_i\theta)^2 + V + V_Q\right) - (\hbar^2/4m)(u^*\partial_{ii}\psi + u\partial_{ii}\psi^*) + \tfrac{1}{2}(u^*\psi + u\psi^*)V. \tag{2.23}$$

As noted, we interpret $J_0(x, q_0, t)$ as the number, or matter, density, so that $\int J_0 d^3x$ is the number of particles in the system. Two points are noteworthy about the expression (2.17): (a) we expect $J_0$ will be non-negative and that in ensuring this the term in brackets may become negative; (b) although $J_0$ exhibits in the first delta-function term the number density of the particle, the right-hand side is not a decomposition into independent particle and field contributions becasue the function $u$ may depend on the particle variables. The latter reamrk applies also to the momentum and energy densities: the right-hand sides of (2.22) and (2.23) are not decompositions into independent particle and field components.

To obtain the corresponding densities for the Schrödinger field alone, we take its Lagrangian density to be

$$\mathcal{L}_S(x) = \tfrac{1}{2}\psi^*\left[i\hbar\dot\psi + (\hbar^2/2m)\partial_{ii}\psi - V\psi\right] + \text{cc}$$
$$= -\rho\left[\dot S + (1/2m)(\partial_i S)^2 + Q + V\right]. \tag{2.24}$$

The function $\psi^*$ ($\psi$) acts like a Lagrange multiplier whose variation implies that the Schrödinger equation for $\psi$ ($\psi^*$) is obtained as a constraint. Formulas (C.5) and (C.6) give the following expressions for the number, number current, momentum and energy densities:

$$j_0 = \rho, \quad j_i = (1/m)\rho\partial_i S \tag{2.25}$$

$$\mathfrak{P}_{Si} = \mathrm{P}/(-\xi_i) = \rho\partial_i S \tag{2.26}$$

$$\mathfrak{H}_S = \mathrm{P}/(-\xi_0) = \rho\left[(1/2m)(\partial_i S)^2 + Q + V\right]. \tag{2.27}$$

Note that $\mathfrak{P}_{Si} = mj_i$ for the pure Schrödinger field whereas $\mathfrak{P} \ne mJ_i$ for the wave-particle composite.

### 3.4 The guidance theorem (single system version)

To ensure that the composite system qualifies for the appellation 'quantum system' our aim is to find the relations linking the functions $\psi$, $q_i$, $V_Q$, $u$ and $\theta$ so that the composite densities (2.17), (2.22) and (2.23) equal their pure Schrödinger counterparts (2.25)-(2.27). These relations must be consistent with the dynamical equations (2.2), (2.3) and (2.15). It is assumed there are no constraints involving $\psi$ alone since the theory applies to arbitrary wavefunctions. We find that the condition of equal densities determines $V_Q$ and $u$ as functions of $\psi$ up to an arbitrary constant, and $\theta$ up to two arbitrary constants:

**Proposition 2 (equality of Noetherian densities)** Suppose the variables $\psi$, $u$ and $\theta$ solve the equations (2.2), (2.3) and (2.15). Then the conserved number, momentum and energy densities coincide with their pure Schrödinger counterparts if and only if the solutions and $V_Q$ are connected by the relations

$$\theta = S - \kappa\log\rho + \text{constant} \tag{2.28}$$

$$V_Q = Q - (\kappa^2/2m)\partial_i\log\rho\,\partial_i\log\rho - (\kappa/m)\partial_{ii}S \tag{2.29}$$

$$u = \psi - \left(1 + \frac{2i\kappa}{\hbar}\right)\frac{1}{\psi^*}\delta(x - q(t)) \tag{2.30}$$

where $\kappa$ is an undetermined constant. (Proof: Appendix D)

As mentioned previously, the functional relations we have found exhaust the constraints obtainable by the method of equating Noetherian densities derived from variational symmetries. Substituting relations (2.28)-(2.30), the composite Lagrangian density (2.16) reduces to the Schrödinger form (2.24), and the number current density (2.18) and the momentum and energy current densities derived from (C.6) all return their Schrödinger values. Conserved densities derived from the number, momentum and energy densities and known functions of $t$ and $x_i$ likewise automatically reduce to their Schrödinger counterparts when the constraints are applied. Examples include the angular momentum density $\varepsilon_{ijk}x_j\mathfrak{P}_k = \varepsilon_{ijk}x_j\mathfrak{P}_{Sk}$ (corresponding to invariance of the action under an infinitesimal rotation $x_i' = x_i - \varepsilon_{ijk}\xi_j x_k$) and the Galilean density $t\mathfrak{P}_i - x_i J_0 = t\mathfrak{P}_{Si} - x_i j_0$ (corresponding to invariance under an infinitesimal boost $x_i' = x_i - \xi_i t$).

To determine the constant $\kappa$ we examine the impact of a time reversal transformation

$$x_i' = x_i, \quad t' = -t, \quad \psi'(x',t') = \psi^*(x,t) \tag{2.31}$$

for which

$$q_i'(t') = q_i(t), \quad \rho'(x',t') = \rho(x,t), \quad S'(x',t') = -S(x,t). \tag{2.32}$$

The corresponding transformations of the fields (2.28)-(2.30) are

$$\theta' = -\theta - 2\kappa\log\rho, \quad V_Q' = V_Q + 2\kappa\partial_{ii}S/m, \quad u' = u^* - (4i\kappa/\hbar\psi)\delta(x - q(t)). \tag{2.33}$$

Assuming the external potential $V$ is a scalar, time reversal is a symmetry of the Schrödinger equation (2.2) but the theory of the composite as a whole is not covariant. The symmetry is broken by the first equation in (2.15), which links the field $\theta$ with the particle velocity ($v_i = m^{-1}\partial_i\theta$). According to the relations (2.31) and (2.32) the velocity of the particle reverses:

$$\dot{q}_i'(t') = \frac{dq_i'(t')}{dt'} = \frac{dq_i(t)}{d(-t)} = -\dot{q}_i(t) \quad \text{or} \quad v_i'(x',t') = -v_i(x,t). \tag{2.34}$$

But, using the first relation in (2.33), we find for the transformed velocity field

$$v_i'(x',t') = m^{-1}\partial_i'\theta' = -v_i(x,t) - 2\kappa m^{-1}\partial_i\log\rho \tag{2.35}$$

which contradicts (2.34) when $\kappa \neq 0$. We conclude that fixing $\kappa$ ($=0$) is tied to requiring time reversal covariance. Consulting (2.33), this may be accomplished by assuming that $V_Q$ is a scalar under the transformation (but evidently alternative assumptions may be invoked, e.g., that the velocity reverses sign):

***Proposition 3 (time reversal)*** Under the conditions of Proposition 2, $V_Q$ is a scalar under time reversal if and only if $\kappa = 0$.

Combining Propositions 2 and 3 we have

***Proposition 4 (full set of constraints)*** The assumptions of Propositions 2 and 3 are obeyed if and only if the functions $\theta$, $u$ and $V_Q$ are fixed as functions of $\psi$ as follows:

$$\theta(x,t) = S(x,t) + \text{constant} \tag{2.36}$$

$$V_Q = Q \tag{2.37}$$

$$u = \psi - \frac{1}{\psi^*}\delta(x - q(t)). \tag{2.38}$$

We note that, given $V_Q = Q$, the condition (2.36) is equivalent to the condition $\partial_i \theta = \partial_i S$, as is easily seen by subtracting (1.4) from (2.15) (and, using Proposition 1 (Sec. 1), (2.36) need only be stated at $t = 0$). Gathering the above results, we obtain the following theorem which states the conditions under which the variables $\dot{q}_i$, $u$ and $V_Q$ are fixed as functions of $\psi$:

***Guidance Theorem 1 (single system version)*** Suppose the variables $\psi$, $u$ and $q_i$ solve the equations (2.2)-(2.4) where the interaction potential $V_Q$ is a function of $\psi$. Let the action with Lagrangian density (2.16) and the function $V_Q$ be invariant, and $u$ transform like $\psi$, with respect to global phase, space and time translations. Then the conserved number, momentum and energy densities (2.17), (2.22) and (2.23) coincide with their quantum counterparts (2.25), (2.26) and (2.27), respectively, and $V_Q$ is a time reversal scalar, if and only if the functions $\psi$, $q_i$, $V_Q$ and $u$ are connected by the relations

$$\dot{q}_i(t) = m^{-1} \partial_i S(x,t)\big|_{x=q(t)} \tag{2.39}$$

$$V_Q = Q \tag{2.40}$$

$$u(x,t,q_0) = \psi(x,t) - \frac{1}{\psi^*(x,t)}\delta(x - q(q_0,t)). \tag{2.41}$$

Thus, the interaction potential is the quantum potential and the Lagrange multiplier $u$ is determined by the wavefunction and the particle coordinates. As anticipated, the only free variables are $q_{0i}$ and $\psi_0$. The dynamical equations governing the wave-particle system therefore become the Schrödinger equation (2.2) and the de Broglie-Bohm guidance equation (2.39) or, following Proposition 1 (Sec. 1),

$$m\ddot{q}_i = -\frac{\partial}{\partial q_i}(V(q) + Q(q)), \quad \dot{q}_i(0) = m^{-1}\partial_i S_0(q_0). \tag{2.42}$$

The Lagrange multiplier obeys the inhomogeneous Schrödinger equation

$$i\hbar \frac{\partial u}{\partial t} = -\frac{\hbar^2}{2m}\partial_{ii}u + V(x)u + 2\frac{\delta Q[\rho(q(t))]}{\delta \psi^*(x,t)} \qquad (2.43)$$

with solution (2.41). The Hamilton-Jacobi equation (2.15) coincides with (1.4). These equations form a mutually consistent set. Equation (2.43) and its solution (2.41) were first given in [8].

So far, the role of the field $u$ has been to assist in building an analytical theory. Its further significance will be examined below. It is easy to evaluate the source term in (2.43) explicitly: using the second of the conversion formulas

$$\frac{\delta}{\delta \psi} = \psi^* \frac{\delta}{\delta \rho} - \frac{i\hbar}{2\psi}\frac{\delta}{\delta S}, \qquad \frac{\delta}{\delta \psi^*} = \psi \frac{\delta}{\delta \rho} + \frac{i\hbar}{2\psi^*}\frac{\delta}{\delta S}, \qquad (2.44)$$

and the following expression for the quantum potential,

$$Q(q) = \frac{\hbar^2}{4m\rho(q)}\left[\left(\frac{1}{2\rho}\frac{\partial \rho(q)}{\partial q_i}\frac{\partial \rho(q)}{\partial q_i} - \frac{\partial^2 \rho(q)}{\partial q_i \partial q_i}\right)\right], \qquad (2.45)$$

the source term is

$$2\frac{\delta Q(q)}{\delta \psi^*(x)} = \frac{\hbar^2}{2m\psi^*(x)}\left[\delta(x-q)\partial_{ii}\log\rho(x) + \partial_i \delta(x-q)\partial_i \log\rho(x) - \partial_{ii}\delta(x-q)\right]. \qquad (2.46)$$

In addition to a term proportional to $\delta(x-q)$ expected in classical particle-source theory, this expression also contains first- and second-order derivatives of the delta function. Note that the factor 2 on the left-hand side of (2.46) is an artefact of the definition of $u$ and can be removed by replacing $u \to 2u$ in (2.16) and (2.41).

### 4 Properties of the wave-particle composite

#### *4.1 Number of particles and mass*

We have interpreted $J_0 = \delta(x-q(t)) + u\psi^*$ in (2.17) as the number density of particles in the system. We expect then that for each $q_{0i}$ the number of particles $\int J_0 d^3x$ will be unity. This is confirmed by inserting $u\psi^* = \rho - \delta(x-q(t))$ in $J_0$. Note that the unity result is due not to the appearance of the particle density (delta-function) in $J_0$, which by construction cancels out, but to the normalized field density $\rho$. In the $n$-body case (see Sec. 8) $J_0$ is the number density of configuration space 'particles' and is again normalized to unity since the $n$ particles make up a single system point.

Hitherto we have not attributed mass to the corpuscle although it is tacit in the Lagrangian (2.1) that the particle has mass and that this is $m$. If we make the latter assumption and bear in mind that the composite system comprises what is usually termed 'a quantum system of mass $m$' (i.e., the $\psi$ field) in addition to the corpuscle, what is the mass of the composite – $2m$? Defining the mass density to be $mJ_0$, an argument similar to that just given for the number density shows that the mass of the composite is, in fact, $m$; by construction the mass density is $m\rho$.

### *4.2 Conservation of energy and momentum*

The self-contained nature of the wave-particle composite may be illustrated by considering the conditions under which its energy and momentum are conserved. Using (2.22) and (2.23) these quantities are given by

$$\int \mathfrak{H} d^3x = \tfrac{1}{2} m \dot{q}_i \dot{q}_i + V(q,t) + V_Q(q(t),t) \\ + \int \left[ -\left(\hbar^2/4m\right)\left(u^* \partial_{ii} \psi + u \partial_{ii} \psi^*\right) + \tfrac{1}{2}\left(u^* \psi + u \psi^*\right) V \right] d^3x \tag{3.1}$$

$$\int \mathfrak{P}_i d^3x = m \dot{q}_i + (i\hbar/2) \int \left(-u^* \partial_i \psi + u \partial_i \psi^*\right) d^3x. \tag{3.2}$$

Differentiating each of these expressions with respect to time and using the dynamical equations (2.2)-(2.4) gives

$$\frac{d}{dt} \int \mathfrak{H} d^3x = \frac{\partial V}{\partial t} + \int u \psi^* \frac{\partial V}{\partial t} d^3x \tag{3.3}$$

$$\frac{d}{dt} \int \mathfrak{P}_i d^3x = -\frac{\partial V}{\partial q_i} - \int u \psi^* \partial_i V d^3x. \tag{3.4}$$

It follows that the energy is conserved when there is no external source of power ($\partial V/\partial t = 0$) and the momentum is conserved when there is no external force ($\partial_i V = 0$). The wave-particle composite as we have defined it is therefore an isolated system when the 'external' agents of change are absent, as expected from the latter's nomenclature. These conditions on the external potential coincide with those under which energy and momentum are conserved according to the usual quantum formalism. Indeed, as we expect, when the solution (2.41) is inserted, (3.3) and (3.4) reduce to the usual quantal expressions for the mean values of the external power and force, respectively:

$$\frac{d}{dt} \int \mathfrak{H}_S d^3x = \int \rho \frac{\partial V}{\partial t} d^3x \tag{3.5}$$

$$\frac{d}{dt} \int \mathfrak{P}_{Si} d^3x = -\int \rho \partial_i V d^3x. \tag{3.6}$$

The energy and momentum of the particle in isolation, whose equations of change are $dE/dt = \partial(V+Q)/\partial t$ and (2.42), respectively, are not conserved unfer these conditions due to the presence of the quantum potential.

### 5 Probability interpretation. Alternative derivation of the guidance equation

The derivation of the de Broglie-Bohm law enshrined in Guidance Theorem 1 (Sec. 3) applies to a single composite system. We thus achieve our aim set out in Sec. 2 of decoupling the law of the individual from a statistical assumption about $\rho$, the latter being regarded in our derivation as a physical field extending throughout space. As remarked in Sec. 2, the statistical assumption is insufficiently stringent in its usual implementation to entail a specific form for the guidance law (we assume that $\rho$ determines the particle probability as a secondary feature).

We now show that the statistical assumption is sufficient to obtain the guidance law if implemented in the context of the assumptions made in Proposition 2 (Sec. 3). This provides an alternative (probabilistic) completion of the derivation of the law of motion in place of the assumption of time reversal covariance of an individual system used in Proposition 3. There is no reason to doubt the validity of time reversal covariance in this context but, proceeding on this alternative basis, this covariance becomes a derived property. In place of Proposition 3 we have

***Proposition 5 (ensemble)*** Under the conditions of Proposition 2, consider an ensemble of wave-particle composite systems for which the wave aspect $\psi$ is identical and the particle position is distributed as $P(x,t) = \rho(x,t)$. Then $\kappa = 0$. (Proof: Appendix E)

In place of Guidance Theorem 1 we may now assert:

***Guidance Theorem 2 (ensemble version)*** In Guidance Theorem 1 replace the condition that $V_Q$ is a time-reversal scalar by the condition that an ensemble of particles is distributed as $\rho(x,t)$ for all $t$. Then the functions $\psi$, $q_i$, $V_Q$ and $u$ are connected by the relations (2.39)-(2.41).

Note that the converse is not true: as shown in Sec. 2, the de Broglie-Bohm law does not imply $P(x,t) = \rho(x,t)$ for all $t$ unless this condition holds at one instant.

### 6 Unification of the Schrödinger and guidance equations

We have seen that, for an interacting wave-particle system, the quantum potential and the guidance condition emerge from the requirements that there is no reciprocal action of the particle on the wave and no empirical discrepancy between the behaviour of the composite and a 'quantum system'. To this end, an auxiliary field was introduced, the Lagrange muliplier $u$, which, as noted above, is determined completely by the variables $q_i$ and $\psi$. In fact, the relations (2.39)-(2.43) that define the theory exhibit a general mutual dependence that suggests the possibility of reversing fundamental roles. To develop this idea, we take $V_Q = Q$ and focus on the close connection between the guidance equation (2.39) and the inhomogeneous equation (2.43) by deriving the latter's general solution:

***Proposition 6 (general solution of the inhomogeneous equation)*** The general solution of (2.43) is

$$u(x,q_0,t) = \phi(x,t) - \frac{1}{\psi^*(x,t)} \delta(x - q(t,q_0))$$
$$+ \int \left[ \dot{q}_i(t') - \frac{1}{m} \frac{\partial S(q(t'),t')}{\partial q_i(t')} \right] \frac{\partial}{\partial q_i(t')} \left[ \frac{G(x - q(t'), t - t')}{\psi^*(q(t'),t')} \right] dt' \quad (5.1)$$

where $\phi$ and $\psi$ obey the homogeneous (Schrödinger) equation and $G(x-x', t-t')$ is the latter's retarded Green function. (Proof: Appendix F)

We see from (5.1) that the legitimacy of the guidance equation (2.39) is correlated with the solution to the inhomogeneous equation (2.43); a deviation in the former is represented in the latter. This result prompts a change of perspective. For it is evident that, while our previous results show that the auxiliary field $u$ is derived from $\psi$ and $q_i$ (Guidance Theorem 1 implies $\phi = \psi$), it may equally be regarded as the theory's primary descriptive element. Treating $\psi$ and $q_i$ as $u$'s constituents, its equation unifies their (one-way) interaction: the homogeneous part is the Schrödinger equation obeyed by the unmodified $\psi$ and, as we show next using the solution (2.41), *the inhomogeneous equation is equivalent to the particle guidance equation*:

***Guidance Theorem 3 (unified version)***. Let a material system of mass *m* be associated with a complex field $u(x,t,q_0)$ that obeys the inhomogeneous Schrödinger equation

$$i\hbar \frac{\partial u(x,t)}{\partial t} = -\frac{\hbar^2}{2m} \partial_{ii} u(x,t) + V(x) u(x,t) + 2 \frac{\delta Q(\rho(q))}{\delta \psi^*(x)}\bigg|_{q=q(t,q_0)} \quad (5.2)$$

where $\psi$ satisfies the homogeneous (Schrödinger) equation, $Q$ is the quantum potential constructed from $\psi$, $q_i(t,q_0)$ are the coordinates of a mobile singularity with initial position $q_{0i}$, and the source term is given in (2.46). Then the function

$$u(x,t,q_0) = \psi(x,t) - \frac{1}{\psi^*(x,t)} \delta(x - q(t,q_0)) \quad (5.3)$$

satisfies (5.2) if and only if the singularity-particle coordinates obey the guidance formula

$$\dot{q}_i = m^{-1} \partial_i S(x)\big|_{x=q(t,q_0)}. \quad \text{(Proof: Appendix G)} \quad (5.4)$$

These results validate the following model: a physical system comprises wave ($\psi$) and particle ($q_i$) aspects, which together produce, via a current generated by the functional gradient of the quantum potential concentrated around the particle trajectory, a 'matter field' described by the amplitude (5.3). The latter integrates the characteristics of both the wave (through the linear component $\psi$) and the particle (through the delta function) into a single spacetime field. The function $u$ may therefore be regarded as an alternative field-theoretic description of the state of the composite system. Employing the solution (5.3), equation (5.2) governing the state melds the Schrödinger equation for the wave (its homogeneous part) with the guidance equation (5.4) (the inhomoeneous equation), the latter being the dynamical law of the delta-function singularity representing the particle. That $\psi$ suffers no reaction from the particle finds an explanation, for it is the field $u$ of which the particle (along with $\psi$) is a source whereas $\psi$ is its (sourceless) homogeneous part. In short: the wave-particle composite both contributes to the source of $u$ and is represented in its structure.

## 7 Properties of the unified model

**(i)** ***Comparison with the double solution***. In our model the source (2.46), built from the wave $\psi$ and particle $q_i$, produces a field $-(1/\psi^*)\delta(x-q)$ that itself has the character of a point particle, namely, a delta function peaked around the particle trajectory (modulated by $1/\psi^*$). The particle is therefore modelled as a mobile singularity moving in accordance with the guidance formula (5.4). Outside the singularity, $u$ coincides with the relatively weak background field $\psi$.

Referring to Appendix B, we have therefore derived a formula of the type (B.1), with $C = 1$. The spreading of $\psi$ implies the spreading of $u$ but this does not undermine the integrity of the singularity whose solitary character is a permanent feature of $u$'s structure. *The particle may be identified with the singular component of the field u.*

Because of the appearance of the inverse complex conjugate wavefunction in the solution $u$, the model also implies *the equality of the phases* of the $u$ and $\psi$ waves in all space, including the region occupied by the particle. Using the polar representation, we have

$$u(x,t) = \left[\sqrt{\rho(x,t)} - \frac{1}{\sqrt{\rho(x,t)}}\delta(x - q(t,q_0))\right]e^{iS(x,t)/\hbar}. \tag{6.1}$$

The particle and the $\psi$ wave thus 'beat in phase' (and hence do not 'interfere'). This provides support for de Broglie's contention that the corpuscle comprises a periodic process locked into the surrounding $\psi$ wave.

We have therefore shown that the ideas informing the pilot-wave and double-solution theories are compatible rather than antagonistic; in our approach the latter is a reformulation of the former where, in particular, the $\psi$ wave retains its dual characteristics of physical field and probability amplitude.

We mention also some points of disparity with de Broglie's version. De Broglie's assumption that the linear component $\psi$ may be multiplied by a constant $C$ as in (B.1) is not borne out since, although the resulting function obeys the inhomogeneous equation, the conditions of Guidance Theorem 1 require $C = 1$. A further difference with de Broglie's approach appears in the extension of our theory to an $n$-body system (Sec. 8) where, in general, a single many-body field $u$ is associated with the system rather than a collection of $n$ 'one-body' fields.

**(ii) *Superposition*.** Given a set of solutions $\psi_\mu$ to the homogeneous equation, the linear superposition $\sum_\mu c_\mu \psi_\mu$, where $c_\mu$ is constant, is also a solution. To each wave $\psi_\mu$ and associated trajectory $q_{\mu i}(t)$ there corresponds a solution $u_\mu$ of the sort (5.3) connected with a quantum potential $Q_\mu(\psi_\mu)$. The solution of the inhomogeneous equation corresponding to the superposition is

$$u(x,q_0,t) = \sum_\mu c_\mu \psi_\mu - \frac{1}{\sum_\mu (c_\mu \psi_\mu)^*}\delta(x - q(t,q_0)) \tag{6.2}$$

where $q_i(t)$ is computed using the total wavefunction, as is $Q$. There is no simple relation between $u$ and the $u_\mu$s, or $Q$ and the $Q_\mu$s. In particular, the combination $u' = \sum_\mu c_\mu u_\mu$ does not generally represent a solution to the inhomogeneous equation. Rather, it obeys an equation involving a sum of source terms, each depending on one of the $\psi_\mu$s, and this sum cannot generally be expressed as a single source depending just on the combination $\sum_\mu c_\mu \psi_\mu$.

**(iii) *Covariance group of the inhomogeneous wave equation*.** Eq. (5.2) is covariant under the continuous transformations of the Galileo group,

$$t' = t + d, \quad x'_i = A_{ij}x_j - w_i t + c_i, \quad q'_i = A_{ij}q_j - w_i t + c_i, \quad \det A = 1, \tag{6.3}$$

where $d, c_i, A_{ij}, w_i$ are constants. This is readily checked on noting that the field (6.1) transforms like $\psi$, that is,

$$u'(x',t') = u(x,t) e^{i\left(\frac{1}{2} m w_i w_i t - m A_{ij} w_i x_j\right)/\hbar}, \tag{6.4}$$

and that the source term (2.46) is $1/\psi^*$ times a scalar. The theory is also covariant with respect to time reversal (for scalar $V$):

$$x'_i = x_i, \quad t' = -t, \quad \psi'(x',t') = \psi^*(x,t), \quad q'_i(t') = q_i(t), \quad u'(x',t') = u^*(x,t). \tag{6.5}$$

**(iv)** *Analogy with field-particle interaction in electromagnetism.* The theory based on the inhomogeneous equation (5.2), where the particle density appears in the source and a general solution $u$ is the superposition of a free solution and a source solution, is analogous, in broad terms, to a typical classical theory of particle-field interactions. However, the details of the quantum version differ significantly because the field $\psi$ and particle enjoy a more intimate relationship than in the classical template, which tends to emphasize their separateness. For definiteness, and with due recognition that we have studied a non-relativistic system, we shall compare with the electromagnetic case where the general solution of the inhomogeneous Maxwell equations with a point-charge source, $\Box A^\mu(x) = e \int \dot{q}^\mu \delta(x - q(\tau)) d\tau$, is the superposition of a free field (obeying the homogeneous wave equation) and the Liénard-Wiechert potentials generated by the particle, and the particle is subject to the Lorentz force law (the 'guidance equation'). Some key differences between the quantum and classical-electromagnetic cases are: (a) the solution $\psi$ to the homogeneous equation is involved in the source (2.46) of the field $u$, which is evident also in the appearance of $\psi$ as a factor of the delta function in (5.3); (b) the quantum source (2.46) is not localized just on the particle track but receives contributions from its neighbourhood through derivatives of the delta function; (c) issues surrounding radiative energy loss, radiation reaction and mass renormalization that are central to the electromagnetic theory [12] are absent here. In particular, by construction, the energy of the quantum composite is that of the homogeneous field; (d) the quantum wave and particle equations are incorporated in the equation for the unified field $u$.

## 8 Many-body systems

For a composite system of $n$ bodies with masses $m_r, r = 1, \ldots, n$, and wavefunction $\psi(x_1, \ldots, x_n, t)$, the function $u$ obeys the inhomogeneous equation

$$i\hbar \frac{\partial u}{\partial t} = -\sum_{r=1}^{n} \frac{\hbar^2}{2m_r} \partial_{ri} \partial_{ri} u + V(x_1, \ldots, x_n) u + 2 \frac{\delta Q(\rho(q_1, \ldots, q_n))}{\delta \psi^*(x_1, \ldots, x_n)} \bigg|_{q_r = q_r(t)} \tag{7.1}$$

where $Q = -\sum_r (\hbar^2 / 2m_r \sqrt{\rho}) \partial_{ri} \partial_{ri} \sqrt{\rho}$. The $n$-body solution corresponding to the single-body formula (5.3) may be written down immediately by extending the index range:

$$u(x_1, \ldots, x_n, q_{10}, \ldots, q_{n0}, t) = \psi(x_1, \ldots, x_n, t) - \frac{1}{\psi^*(x_1, \ldots, x_n, t)} \prod_{r=1}^{n} \delta(x_r - q_r(q_{10}, \ldots, q_{n0}, t)). \tag{7.2}$$

Following Guidance Theorem 3, this solution renders the inhomogeneous equation (7.1) equivalent to the many-body guidance equation

$$\dot{q}_{ri} = \frac{1}{m_r} \partial_{ri} S(x_1,...,x_n)\Big|_{x_r = q_r(q_{r0},t)}, \quad r = 1,...,n. \quad (7.3)$$

The function $u$ is therefore defined irreducibly in the configuration spaces spanned by $x_{1i},...,x_{ni}$ and $q_{10i},...,q_{n0i}$. On the other hand, the single configuration space trajectory may be regarded as composed of $n$ three-dimensional trajectories $q_{ri}(q_{10},...,q_{n0},t)$, where each triplet of coordinates ($i = 1,2,3$) for given $r = 1,..., n$ corresponds to one of the $n$ corpuscles making up the particle component of the system. Each particle's coordinates generally depend on the labels of the other $n$-1 particles, which expresses the nonlocal connection of the set.

When the wavefunction factorizes into $n$ one-body factors, $\psi(x_1,...,x_n) = \prod_{r=1}^n \psi_r(x_r)$, the particle delta function does likewise, and the system decomposes into a set of $n$ independent 'single-body' composite systems. The unified field $u$ can be expressed in terms of the single-body fields $u_r$ as a sort of factorization:

$$u - \psi = \prod_{r=1}^n [u_r(x_r, q_{r0}) - \psi_r(x_r)]. \quad (7.4)$$

Using the function $u$ in (7.4), (7.1) becomes equivalent to a set of $n$ copies of the one-body guidance equation.

## 9 Conclusion

### *9.1 Revised postulates of the causal interpretation*

Our initial aim was to find a non-statistical vindication of the de Broglie-Bohm law that incorporates the non-reactive character of the particle on the $\psi$ wave. To this end, we developed an analytical approach to the dynamics of a single system whose interacting wave and particle components, inseparable yet independent entities according to the usual de Broglie-Bohm theory, are treated as a unit insofar as key properties of the composite - conserved densities implied by assumed variational symmetries - coincide with those of a usual 'quantum system'. In conjunction with the condition of time reversal covariance, the particle equation and interaction potential become the de Broglie-Bohm law and the quantum potential, respectively. We also showed how the time reversal assumption may be replaced by a statistical condition.

In the process, an alternative perspective for the causal theory emerged based on a function $u$, introduced initially as a Lagrange multiplier, for which the particle (together with $\psi$) is a source and whose governing inhomogeneous equation embraces both the de Broglie-Bohm law for the particle and the Schrödinger equation for the wave. The disparate elements of the de Broglie-Bohm theory now become aspects of the single field $u$, with the linear wave being its sourceless homogeneous part and the particle being represented by a highly concentrated amplitude that moves in accordance with the guidance law. The 'true' matter field of quantum theory is then to be identified not with $\psi$ but with $u$, and we have argued that the model provides a realization of de Broglie's hitherto unfulfilled double solution programme. Since the representation of the particle by the $u$ field is a (peaked) point following the track implied by the guidance law, its spatial probability distribution is identical to that of the particle postulated in the usual de Broglie-Bohm theory.

We summarize these findings in the following set of revised postulates for the causal theory:

1′ A quantum system comprises a wave-particle composite whose state is described by a field $u$ that obeys the inhomogeneous equation (5.2).

2′ The field solution is $u(x,t,q_0) = \psi(x,t) - (1/\psi^*(x,t))\delta(x - q(t,q_0))$ where the delta function represents the particle aspect.

3′ For an ensemble of $u$ fields with a common $\psi$ component and an initial particle amplitude $\delta(x - q_0)$ whose location varies with $q_{0i}$, the initial probability density is $\rho_0(q_0)$.

The justification for postulate 2′ is that of postulate 2 (see Guidance Theorem 1, Sec. 3). We have shown in Guidance Theorem 3 (Sec. 6) that the combination of postulates 1′ and 2′ furnishes the guidance law for the particle amplitude and this law is selected (within the analytical scheme we have used) as that which secures empirical equivalence with quantum theory. The extension of the postulates to many-body systems is straightforward, using the results of Sec. 8. The unified theory may be applied in an obvious way simply by replacing the point particle of the usual de Broglie-Bohm theory by the microscopic particle amplitude. It therefore provides a satisfactory alternative causal underpinning for quantum mechanics.

### 9.2 Open questions

In our analytical scheme we demonstrated the constructive role played by spacetime (translation) and internal (gauge) symmetries in selecting dynamical equations. This was achieved through conditions imposed on the conserved densities generated by the symmetries via Noether's theorem. The latter represents a potential limitation of the method; the correlation between a symmetry and a density established thereby is not unique since it depends intimately on the Lagrangian used in deriving the Euler-Lagrange equations. The latitude in the Lagrangian density goes considerably beyond the addition of a total divergence, and quite diverse expressions for a conserved density may accompany a given symmetry. Thus, while the analytical method detaches the guidance formula from the statistical postulate, the result may not be unique. Likewise, the inhomogeneous equation may be open to modification although it should be noted that it's property of unifying the Schrödinger and guidance equations is independent of the analytical procedure that produced it. Further justification of this equation may also focus on the physical basis of the absence of particle back-reaction.

In examining alternative techniques to justify the revised postulates, we may consider utilizing a relativistic treatment since Lorentz symmetry is known to enforce a unique expression for the particle law in certain contexts [7]. We might also base the theory directly on an analysis of the forces acting within the system, and entertain less restrictive assumptions, such as allowing a more general interaction than a scalar potential or imbuing the particle with structure. In relation to the latter, the hydrodynamic analogy prompts the notion that the inertia of the particle may be acquired through a 'virtual mass' effect stemming from the displacement of the enveloping fluid [13]. We may also relax the requirement that the model reproduces exactly the current empirical content of quantum mechanics although there are as yet no clues as to where deviations might occur. A further issue is the impact on the unified theory of employing the trajectory conception of the quantum state in place of the wavefunction [4,5,14]. This is a kind of 'prequel' to the de Broglie-Bohm theory in that the entire fleet of potential trajectories is utilized to define the quantum state (from which the time-dependence of the wavefunction may be derived), but shorn of the additional corpuscle. This approach provides a congenial setting to introduce the latter.

## Appendices

### A Proof of Proposition 1

Taking the gradient of (1.4) and rearranging we obtain

$$\left(\frac{\partial}{\partial t}+\frac{1}{m}\partial_j S\partial_j\right)\partial_i S=-\partial_i(V+Q). \tag{A.1}$$

Passing to the comoving coordinates $x_i = q_i(t)$ and employing the first-order law (1.2), (A.1) may be written as the second-order equation

$$m\ddot{q}_i = -\partial_i(V+Q)\big|_{x=q(t)}. \tag{A.2}$$

Conversely, to obtain the first-order law from the second-order one, we evaluate the functions in (A.1) along the trajectory and subtract from (A.2) to get

$$\frac{d}{dt}\left(m\dot{q}_i - \partial_i S\big|_{x=q(t)}\right) = -\left(m\dot{q}_j - \partial_j S\big|_{x=q(t)}\right)\partial_{ij}S\big|_{x=q(t)}. \tag{A.3}$$

This is a first-order linear ordinary differential equation $\dot{X}_i = A_{ij}(t)X_j(t)$ for which continuity of the matrix $A_{ij}(t)$ guarantees the existence and uniqueness of solutions $X_i(t)$ [15]. Then, since $m\dot{q}_i = \partial_i S$ is a solution of (A.3), this is the unique solution for all $t$ if it holds at $t = 0$, granted the continuity of the functions $\partial_{ij}S$.

### B De Broglie's objections to the pilot wave. The double solution

The most comprehensive single source for de Broglie's version of the pilot-wave theory – including an account of his 1920s work, his response to Bohm's 1952 reworking, his critique of the theory, and his remedy for what he saw as its deficiencies (the 'double solution') – is the second part of his 1956 book translated as '*Non-Linear Wave Mechanics*' [16]. Unless otherwise stated the views attributed to de Broglie here are drawn from this source (de Broglie later added further speculations regarding, in particular, the quantum potential (e.g., [17]) but he made no further substantive advances in the topics discussed here). As is evident from the book's title, de Broglie sought an alternative to the pilot wave's rendering of linear quantum theory. Actually, one has to work hard to extract a straightforward account of the pilot-wave theory from de Broglie's expositions as they were often intermingled with criticisms of the theory, the 'double solution' alternative, and extraneous and contentious material on relativistic formulations. Much of the latter material, involving the Klein-Gordon equation, 'photon trajectories' and an imaginary variable mass, is inconsistent and gives the misleading impression that the basic ideas require relativity. In fact, although de Broglie was an architect of the nascent pilot-wave theory in the 1920s, he was a critic of it from the outset. He believed the theory is basically inconsistent and at best a fragmentary and provisional view, to be superseded eventually by his cherished 'double solution' proposal. His criticisms revolve around two broad themes: the relation between the wave and the particle, and the nature of the wavefunction. Regarding the latter, his disquiet unfortunately concerned not the problematic relativistic material just mentioned but features that are either unproblematic or are among the pilot-wave theory's more valuable insights. Moreover, we will see that his views are not always consistent.

With reference to the $\psi$ wave, de Broglie believed that it is 'fictitious' and hence cannot play the role of a physical field causing observable alterations in a particle's motion, as required by the guidance theory. That is, in de Broglie's view, the 'matter wave' he famously introduced in 1923 as an objective attribute of material systems is *not* to be identified with Schrödinger's wavefunction of 1926. He deemed the wavefunction to be fictional in two respects. First, he felt that $\psi$, being a repository of statistical knowledge about the location of a particle, cannot be simultaneously descriptive of a real physical situation. In support of this contention he appealed to the authority of classical physics where one finds an absolute distinction between fields that are either physical or fictitious (probability distributions); no single entity combines both features. Second, for a many-body system where $\psi$ is generally defined irreducibly in the configuration space of all the particles, de Broglie judged this arena 'obviously fictitious' as it could not be identified with or mapped into three-dimensional space. He believed that, while configuration space is a facet of classical physics (as in the Hamilton-Jacobi theory, for example), it is used as a matter of convenience and, if desired, one can translate the multidimensional theory into the framework of three-dimensional space, something he suggested is generally impossible with the $\psi$ function. Hence, he argued, for these two reasons a theory such as the pilot wave, that is based on attributing physical effects to $\psi$, is basically inconsistent.

De Broglie appears to be the first to identify the quantum potential as the causal agent in the pilot-wave theory, invoking the 'force of a new kind' [18] that it generates as the explanation for the deviation in particle motion, especially in interference effects. But he felt that the special properties of the quantum potential are additional factors weakening the appeal of the theory. Noting that the quantum potential depends on the form of the wave rather than its absolute magnitude, he expressed misgivings that this potential could have a significant effect on its associated particle in the case of a wave of small amplitude, since classically one would expect the wave's effect would be proportional to the amplitude and hence, in the case considered, negligible. He also commented adversely on the intervention of the quantum potential in the particle energy and momentum conservation laws [19].

It is notable that de Broglie's objections to the pilot wave turn upon unfavourable comparisons with pre-quantum theories of particles and fields. He could not countenance the step away from 'classical conceptions' demanded by the full deployment of his pilot-wave theory; he abjured its essential nonclassicality in favour of some kind of local three-dimensional theory of waves and particles. The essence of de Broglie's objections may therefore be summarized thus: *the pilot-wave theory is not a classical theory*. The remit of his double solution programme was then to produce a theory more in accord with 'classical conceptions'. De Broglie wanted this to be not a new dynamics but an old dynamics, modelled on the nineteenth century methods of Hamilton and Jacobi and the classical unified field theory advocated by Einstein, yet somehow realized in a quantum context.

It must be said that de Broglie's case for rejecting the pilot wave, which forms part of the justification for introducing the double solution, is rather flimsy. It resembles the kind of case made by defenders of alternative interpretations who believe the pilot-wave theory may be dismissed on the basis of a few superficial observations, and it is easily countered [2]. Attributing a physical status to a statistical function ($\psi$) is not a logical problem; one simply reverses the roles and asserts that the primary property of $\psi$ is that it defines a physical field while, as a secondary property, its amplitude squared is numerically equal to a probability density. This combination of roles may not be a feature of classical fields and one may enquire further as to how $\psi$ comes to have this property but it is not logically objectionable.

De Broglie's concern over ascribing physical status to configuration space was better founded and continues to be debated today, but again this feature does not challenge the consistency of the pilot-wave theory. In any case, his disquiet is ameliorated by consideration of the trajectory conception of the state, according to which the many-body quantum state may be expressed as a set of states in three-dimensional space, one associated with each particle (in a way that is consistent with quantum nonlocality) [20].

Finally, as regards the quantum potential, its possible potency in domains of low amplitude may be unprecedented in classical field theory (although not entirely unknown since fluid potentials may have this property) but novelty should not be a basis for automatic rejection. Its role in the conservation laws is considered in Sec. 4.2.

De Broglie's other broad area of demurral, and his starting point in seeking to transcend the pilot wave, stemmed from his conviction that the physical expression of wave-particle duality could not be achieved by maintaining 'wave' and 'particle' as separate categories whose natures are external to one another, a signature of the usual pilot-wave approach. Rather, he proposed that the wave and particle should be incorporated into a single physical entity, a continuous complex field $u$, that simultaneously displays the dual aspects of an almost-everywhere regular wave and a singular (highly concentrated) persistent mobile region representing the particle. The regular and singular regions would be connected in such a way that the guidance of the latter would be explained by the overall physical structure rather than simply postulated. Specifically, de Broglie hoped to explain the guidance law by supposing that the particle is akin to a clock whose phase is locked to the surrounding regular wave.

There would then be two fundamental fields in quantum physics: the probabilistically-interpreted non-physical $\psi$ wave and the physical field $u$. In the initial (1927) formulation of the idea, de Broglie expected that both functions, $\psi$ and $u$, would obey the same (linear) wave equation; they would thus constitute a 'double solution' of that equation, an appellation he maintained when he proposed in the 1950s that the field $u$ containing the singularity should obey a nonlinear equation while $\psi$ continues to satisfy the linear Schrödinger equation.

De Broglie never gave an example of the purported nonlinear equation for $u$ and his discussion is not entirely consistent. For example, he envisaged that the regular segment of $u$ conveys information about the potentials it encounters to the embedded singularity, much as $\psi$ steers the corpuscle in the pilot-wave theory. To ensure that the singularity obeys the guidance formula (1.2) he therefore suggested concordance of the phases of $u$ and $\psi$, his specific proposal being that $u$ should generally have the form

$$u(x,t) = C\psi(x,t) + \tilde{u}(x,t) \tag{B.1}$$

where $\tilde{u}$ describes an intense region of field representing the corpuscle. De Broglie claimed that the constant factor $C$ resolves the contradiction implicit in (B.1) between the purely subjective role he attributed to the function $\psi$ and the objective function $u$ [21]. However, it is not clear that the model is intelligible unless one supposes $\psi$ refers to a physical field, especially since one may choose $C = 1$ (a condition derived in our approach). The proposal was in any case speculative; de Broglie was unsure whether the equality of phases should extend into the singular region or whether $u$ should coincide with $C\psi$ outside that region since, in view of the spreading of $\psi$ under Schrödinger evolution, this might undermine the solitonic aspect of $u$. To circumvent the 'obviously fictitious' many-body configuration space, de Broglie proposed that the system should comprise a set of three-dimensional $u$ waves, one corresponding to each particle, alongside the multidimensional $\psi$ wave. Again, it is not clear how that notion could be compatible with a formula of the type (B.1). Curiously, given his penchant for 'classical conceptions', and that (B.1) contains the linear component $\psi$, de Broglie does not appear to have considered the possibility of an inhomogeneous linear equation for $u$, similar to that employed in classical theories of field-particle interaction (and as we derive in the text).

De Broglie's objections to the pilot wave do not justify the step of introducing a novel physical field, for which there was (and is today) no empirical authority, and his account was not always coherent. Nevertheless, we show in the text that the programme of linking the wave and the particle at a basic level is justified if recontextualized, for a field of the type (B.1) emerges when the pilot-wave theory is embedded in an analytical framework. This is prompted by issues (described in Sec. 2) not mentioned by de Broglie.

## C Noether's theorem

Suppose $F$ is a functional of some function $\phi(x)$: $F[\phi] = \int f(x,\phi,\partial\phi,\partial^2\phi,...)d^3x$. Then the functional derivative of $F$ with respect to $\phi$ is

$$\frac{\delta F}{\delta \phi} = \frac{\partial f}{\partial \phi} - \partial_i \frac{\partial f}{\partial(\partial_i \phi)} + \partial_{ij} \frac{\partial f}{\partial(\partial_{ij} \phi)} - ... \tag{C.1}$$

If $F$ is a local function of $\phi$ and its derivatives (as $V_Q$ is of $\psi$), so that $F = F(y,\phi(y),\partial\phi(y),\partial^2\phi(y),...)$, then $f = F(x)\delta(x-y)$ and

$$\frac{\delta F(y)}{\delta \phi(x)} = \frac{\partial F(x)}{\partial \phi(x)}\delta(x-y) - \partial_i\left[\frac{\partial F(x)}{\partial(\partial_i \phi)}\delta(x-y)\right] + \partial_{ij}\left[\frac{\partial F(x)}{\partial(\partial_{ij} \phi)}\delta(x-y)\right] - ... \tag{C.2}$$

Let a physical system be described by fields $X_\mu(x,t)$ and suppose the Lagrangian density depends on time derivatives of at most first order and space derivatives of any order: $\mathcal{L}(X,\dot{X},\partial X,\partial^2 X,...,x,t)$ ((2.16) is an example). Let the action $\int \mathcal{L} d^3x\,dt$ be invariant under the following infinitesimal transformation of the independent and dependent variables:

$$t' = t + \varepsilon\xi_0(x,t), \quad x'_i = x_i + \varepsilon\xi_i(x,t), \quad X'_\mu(x',t') = X_\mu(x,t) + \varepsilon\eta_\mu(x,t). \tag{C.3}$$

Noether's (first) theorem [22] asserts that, when the fields satisfy the Euler-Lagrange equations

$$\frac{\partial}{\partial t}\frac{\partial \mathcal{L}}{\partial \dot{X}_\mu} = \frac{\delta L}{\delta X_\mu} \equiv \frac{\partial \mathcal{L}}{\partial X_\mu} - \partial_i \frac{\partial \mathcal{L}}{\partial(\partial_i X_\mu)} + \partial_{ij} \frac{\partial \mathcal{L}}{\partial(\partial_{ij} X_\mu)} - ..., \tag{C.4}$$

the following scalar and vector expressions,

$$\mathrm{P} = \mathcal{L}\xi_0 + \frac{\partial \mathcal{L}}{\partial \dot{X}_\mu}\left(\eta_\mu - \xi_0\dot{X}_\mu - \xi_l\partial_l X_\mu\right) - \Lambda_0 \tag{C.5}$$

$$\begin{aligned}P_j = \mathcal{L}\xi_j &+ \left(\frac{\partial \mathcal{L}}{\partial(\partial_j X_\mu)} - \frac{D}{\partial x_k}\frac{\partial \mathcal{L}}{\partial(\partial_{jk} X_\mu)} + \frac{D^2}{\partial x_k \partial x_m}\frac{\partial \mathcal{L}}{\partial(\partial_{jkm} X_\mu)} - \ldots\right)\left(\eta_\mu - \xi_0 \dot{X}_\mu - \xi_l \partial_l X_\mu\right) \\ &+ \left(\frac{\partial \mathcal{L}}{\partial(\partial_{jk} X_\mu)} - \frac{D}{\partial x_m}\frac{\partial \mathcal{L}}{\partial(\partial_{jkm} X_\mu)} + \ldots\right)\frac{D}{\partial x_k}\left(\eta_\mu - \xi_0 \dot{X}_\mu - \xi_l \partial_l X_\mu\right) + \ldots - \Lambda_j, \end{aligned}$$ (C.6)

together obey the continuity eqaution

$$\frac{\partial \mathrm{P}}{\partial t} + \partial_i \mathrm{P}_i = 0$$ (C.7)

for some functions $\Lambda_0, \Lambda_i$. The notation '$D$' means differentiate with respect to all functions of $x_i$. For a given Lagrangian density, (C.7) determines the variational symmetries, i.e., the functions $\xi_0, \xi_i, \eta_\mu$ which characterize the invariance transformations of the action, along with the functions $\Lambda_0, \Lambda_i$ and conditions on the external potentials. Then, to each variational symmetry (which is also a symmetry of the Euler-Lagrange equations) there corresponds a local density (C.5) and consequently a global charge $\int P(x,t) d^3 x$, which from (C.7) is conserved assuming suitable boundary conditions on the fields. Note that the functions $\mathrm{P}, \mathrm{P}_i$ inherit the parameters contained in the functions $\xi_0, \xi_i, \eta_\mu$ but the arbitrariness of the parameters means we may extract conserved quantities independent of them. The local quantities $\mathrm{P}$ may have direct physical significance as well as their global values and, in the absence of constraints on $\mathrm{P}$, a given charge corresponds to an infinite set of densities.

Instead of deriving the symmetries from (C.7) according to the method just outlined, in the text we follow the more common inverse procedure: we notice, or require, that certain transformations leave the action invariant and write down the associated conserved densities using (C.5) (tacitly assuming the corresponding conditions on the external potentials where necessary). In our application the functions $\Lambda_0 = \Lambda_i = 0$ in all cases.

## D Proof of Proposition 2

It is convenient to replace $u$ in the formulas for the conserved densities by the function $\tilde{u}$ defined by

$$\tilde{u} = u - \psi + \frac{1}{\psi^*}\delta(x - q(t)).$$ (D.1)

Equating the densities (2.17), (2.22) and (2.23) with (2.25), (2.26) and (2.27), respectively, then gives the relations

$$\tilde{u}^*\psi + \tilde{u}\psi^* = 0$$ (D.2)

$$\partial_i(\theta - S)\delta(x - q) = -(i\hbar/2)\left(-\tilde{u}^* \partial_i \psi + \tilde{u} \partial_i \psi^*\right)$$ (D.3)

$$\begin{aligned}\left[(1/2m)(\partial_i\theta\partial_i\theta-\partial_iS\partial_iS)+V_Q-Q\right]\delta(x-q)\\ =\left(\hbar^2/4m\right)(\tilde{u}*\partial_{ii}\psi+\tilde{u}\partial_{ii}\psi*)-\tfrac{1}{2}(\tilde{u}*\psi+\tilde{u}\psi*)V.\end{aligned} \qquad (D.4)$$

Using (D.2) we can eliminate $\tilde{u}*$ in (D.3) and (D.4) to get the four equations

$$\partial_i(S-\theta)\delta(x-q)=(i\hbar/2\psi)\tilde{u}\partial_i\rho \qquad (D.5)$$

$$\left[(1/2m)(\partial_i\theta\partial_i\theta-\partial_iS\partial_iS)+V_Q-Q\right]\delta(x-q)=-(i\hbar/2m\psi)\tilde{u}\partial_i(\rho\partial_iS). \qquad (D.6)$$

It follows from (D.5) that $(i\hbar/2\psi)\tilde{u}=\mu(x)\delta(x-q)$ and so $\partial_i\theta=\partial_iS-\mu\partial_i\rho$ for some real function $\mu(x)$ that is to be determined. Since $\mu\partial_i\rho$ is a gradient field, $\mu$ is a function just of $\rho$ so that $\mu\partial_i\rho=\partial_i f(\rho)$ with $\mu=df/d\rho$. Using this result in (D.6) gives an expression for $V_Q$ in terms of $\rho$ and $S$. In sum, we have the following necessary conditions for equality of each of the two species of number, momentum and energy density, respectively:

$$\partial_i\theta=\partial_i(S-f(\rho)) \qquad (D.7)$$

$$u=\psi-\left(\frac{1}{\psi*}+\frac{2i\psi\mu}{\hbar}\right)\delta(x-q(t)) \qquad (D.8)$$

$$V_Q=Q-(1/2m)\partial_i f\partial_i f-(1/m)\mu\rho\partial_{ii}S. \qquad (D.9)$$

These results enable us to determine $\theta$ in terms of $\rho$ and $S$. From (D.7) we have $\theta(x,t)=S(x,t)-f(\rho(x,t))+F(t)$. To find $F$, we substitute this expression in the Hamilton-Jacobi equation (2.15). Subtracting (1.4) and inserting the expression (D.9) leaves $\dot{F}=0$, that is, $F$ = constant. Note that we used here the particle equation (2.15) and hence the second-order version (2.4) does not supply further restrictions.

To fix $\mu$ further, it remains to take account of the requirement that $u$ obeys (2.3). Substituting (D.8) and (D.9) in (2.3) and using (2.2), the $u$ equation reduces to

$$\begin{aligned}\left(2\mu\psi-\frac{i\hbar}{\psi*}\right)\left\{\frac{\partial}{\partial t}\delta(x-q(t))+m^{-1}\partial_i\left[\delta(x-q(t))\partial_i(S-f(\rho))\right]\right\}\\ =\frac{2\psi}{m}\delta(x-q(t))\partial_{ii}S(\mu'\rho+\mu).\end{aligned} \qquad (D.10)$$

Using (2.7), the expression in curly brackets on the left-hand side is identically zero. Then from the right-hand side $\mu'\rho+\mu=0$ whence $\mu=\kappa/\rho$ where $\kappa$ = constant. Substituting in (D.7)-(D.9) then gives (2.28)-(2.30). Conversely, it is easy to check that the conditions (2.28)-(2.30) are sufficient to ensure equality of the Schrödinger and composite number, momentum and energy densities.

# E Proof of Proposition 5

Let the initial particle density be denoted $P_0(q_0)$. Multiplying the identity (2.13) by $P_0$ and integrating over $q_{0i}$, we obtain the continuity equation

$$\frac{\partial P}{\partial t} + \partial_i(P v_i) = 0 \tag{E.1}$$

where

$$P(x,t) = \int P_0(q_0) \delta(x - q(q_0,t)) d^3 q_0 \tag{E.2}$$

$$P(x,t) v_i(x,t) = \int P_0(q_0) \dot{q}_i(q_0,t) \delta(x - q(q_0,t)) d^3 q_0 \tag{E.3}$$

with $v_i(x,t) = m^{-1} \partial_i(S - \kappa \log \rho)$ from (2.28). Setting $P(x,t) = \rho(x,t)$ and subtracting (1.3) from (E.1) implies $\kappa \nabla^2 \rho = 0$. Hence $\kappa = 0$ or $\rho$ is a harmonic function. For the boundary condition $\rho \to 0$ as $x_i \to \pm\infty$ the unique solution to Laplace's equation is $\rho = 0$ for all $x_i$. Hence $\kappa = 0$.

# F Proof of Proposition 6

The retarded Green function satisfies [26]

$$\left( i\hbar \frac{\partial}{\partial t} + \frac{\hbar^2}{2m} \partial_{ii} - V(x,t) \right) G(x - x', t - t') = i\hbar \delta(x - x') \delta(t - t') \tag{F.1}$$

with

$$\Theta(t - t') \psi(x,t) = \int G(x - x', t - t') \psi(x',t') d^3 x' \tag{F.2}$$

where $\Theta(t > 0) = 1$ and $\Theta(t < 0) = 0$. The general solution of (2.43) is the superposition of the complementary function ($\phi$, obeying the Schrödinger equation) and the particular integral $u_p$: $u = \phi + u_p$ (in the general case the function $\phi$ may differ from the homogeneous solution $\psi$ used in the source). The particular integral is readily checked by substitution in (2.43) to be

$$u_p(x,t) = \frac{2}{i\hbar} \int_{-\infty}^{\infty} G(x - x', t - t') \frac{\delta Q(\rho(q(t')))}{\delta \psi^*(x',t')} d^3 x' dt'. \tag{F.3}$$

Inserting the source (2.46) this becomes

$$u_p(x,t) = \frac{\hbar}{2mi} \int \frac{G(x - x', t - t')}{\psi^*(x',t')} \left[ \delta(x' - q(t')) \partial'_{ii} \log \rho(x',t') \right. \\
\left. + \partial'_i \delta(x' - q(t')) \partial'_i \log \rho(x',t') - \partial'_{ii} \delta(x' - q(t')) \right] d^3 x' dt'. \tag{F.4}$$

Writing $\delta = \delta(x' - q(t'))$ and using the properties $\partial'_i \delta = -\partial \delta / \partial q_i$ and $\partial'_{ii} \delta = \partial^2 \delta / \partial q_i \partial q_i$, we take the $q_i$ derivatives out of the integral, perform the $x_i$ integration, differentiate the integrand with respect to $q_i$, and reinstate $\delta$. This gives

$$u_p(x,t) = \int \left[ -\frac{1}{m} \partial'_i S \partial'_i \left( \frac{G}{\psi^*} \right) + \frac{\hbar}{2mi\psi^*} \left( \frac{G}{\psi^*} \partial'_{ii} \psi^* - \partial'_{ii} G \right) \right] \delta(x' - q(t')) d^3x' dt'. \tag{F.5}$$

Noting that, for the primed variables, $G$ obeys

$$\left( -i\hbar \frac{\partial}{\partial t'} + \frac{\hbar^2}{2m} \partial'_{ii} - V(x',t') \right) G(x-x', t-t') = i\hbar \delta(x-x') \delta(t-t'), \tag{F.6}$$

and substituting for $\partial'_{ii} G$ from this relation and $\partial'_{ii} \psi^*$ from the complex conjugate Schrödinger equation, (F.5) becomes

$$u_p(x,t) = -\frac{1}{\psi^*(x,t)} \delta(x - q(t)) - \int \left[ \frac{\partial}{\partial t'} \left( \frac{G}{\psi^*} \right) + \frac{1}{m} \partial'_i S \partial'_i \left( \frac{G}{\psi^*} \right) \right] \delta(x' - q(t')) d^3x' dt'. \tag{F.7}$$

Writing

$$\delta(x' - q) \frac{\partial}{\partial t'} \left( \frac{G}{\psi^*} \right) = \frac{d}{dt'} \left( \frac{G}{\psi^*} \delta(x' - q) \right) - \frac{G}{\psi^*} \dot{q}_i(t') \frac{\partial}{\partial q_i(t')} \delta(x' - q) \tag{F.8}$$

where $d/dt' = \partial/\partial t' + \dot{q}_i(t') \partial / \partial q_i(t')$ and completing the $x_i$ integration gives

$$u_p(x,t) = -\frac{1}{\psi^*(x,t)} \delta(x - q(t)) - \int \frac{d}{dt'} \left[ \frac{G(x - q(t'), t - t')}{\psi^*(q(t'), t')} \right] dt'$$
$$+ \int \left[ \dot{q}_i(t') - \frac{1}{m} \frac{\partial S(q(t'), t')}{\partial q_i(t')} \right] \frac{\partial}{\partial q_i(t')} \left[ \frac{G(x - q(t'), t - t')}{\psi^*(q(t'), t')} \right] dt'. \tag{F.9}$$

It follows from (F.1) that the first integral in (F.9) obeys the homogeneous equation and so this term can be absorbed in the arbitrary homogeneous function $\phi$. Eq. (5.1) follows.

### G Proof of Guidance Theorem 3

We shall give two proofs, the first based on Proposition 6. Assuming (5.3), (5.1) (with $\phi = \psi$) implies

$$\int \left[ \left( \dot{q}_i(t') - \frac{1}{m} \frac{\partial S(q(t'), t')}{\partial q_i(t')} \right) \frac{\partial}{\partial q_i(t')} \left( \frac{G(x - q(t'), t - t')}{\psi^*(q(t'), t')} \right) \right] dt' = 0. \tag{G.1}$$

Differentiating (G.1) with respect to $t$ and $x_i$ and using (F.1), we obtain

$$\left[\dot{q}_i(t) - \frac{1}{m}\frac{\partial S(q(t))}{\partial q_i(t)}\right]\frac{\partial}{\partial q_i(t)}\left[\frac{\delta(x-q(t))}{\psi^*(q(t))}\right] = 0. \tag{G.2}$$

Using the property $f(x)\delta(x-q) = f(q)\delta(x-q)$ this gives

$$\left[\dot{q}_i - \frac{1}{m}\frac{\partial S(q(t))}{\partial q_i(t)}\right]\partial_i\left[\delta(x-q(t))\right] = 0. \tag{G.3}$$

Multiplying by an arbitrary function $f(x)$ and integrating over all $x_i$ implies $(\dot{q}_i - m^{-1}\partial S/\partial q_i)\partial f(q)/\partial q_i = 0$ and hence (5.4) follows. The converse is obvious: inserting (5.4) in (5.1) implies (5.3).

For the second proof we substitute (5.3) directly in (5.2). The latter reduces to a continuity equation for the microscopic particle density (cf. (D.10) with $\mu = 0$):

$$\frac{\partial}{\partial t}\delta(x-q(t)) + m^{-1}\partial_i\left[\delta(x-q(t))\partial_i S(x)\right] = 0. \tag{G.4}$$

Subtracting this from the identity (2.13) implies (G.3) and we proceed as in the first proof. Conversely, the guidance equation (5.4) together with the identity (2.13) implies (G.4) from which we can deduce the inhomogeneous equation (5.2) with the solution (5.3), assuming $\psi$ obeys the Schrödinger equation.

*Acknowledgements: The author thanks Chris Philippidis for valuable comments and an anonymous referee for helpful suggestions.*